\documentclass{aa}
\usepackage{natbib,graphicx,hyperref} 
\usepackage[varg]{txfonts}

\newcommand{\added}{}   
\begin{document}

\title{The turbulent diffusion of toroidal magnetic flux
      as inferred from properties of the sunspot butterfly diagram}

\author{R.H. Cameron\inst{1} and M. Sch{\"u}ssler\inst{1}}

\institute{Max-Planck-Institut f{\"u}r Sonnensystemforschung, 
           Justus-von-Liebig-Weg 3, 37077 G{\"o}ttingen, Germany}
\date{Received ; accepted}

\abstract {In order to match observed properties of the solar
 cycle, flux-transport dynamo models require the toroidal magnetic flux
 to be stored in a region of low magnetic diffusivity, typically located
 at or below the bottom of the convection zone.}
{We infer the turbulent magnetic diffusivity affecting the toroidal
 field on the basis of empirical data.}
{We consider the time evolution of mean latitude and width of the
 activity belts of solar cycles 12--23 and their dependence on cycle
 strength. We interpret the decline phase of the cycles as a diffusion
 process.}
{The activity level of a given cycle begins to decline when the
 centers of its equatorward propagating activity belts come within their
 width (at half maximum) from the equator. This happens earlier for
 stronger cycles because their activity belts are wider. From that 
 moment on, the activity and the belt width decrease in the same
 manner for all cycles, independent of their maximum activity
 level.  In terms of diffusive cancellation of opposite-polarity
 toroidal flux across the equator, we infer the turbulent diffusivity
 experienced by the toroidal field, wherever it is located, to be in the
 range 150--450~km$^2$s$^{-1}$. Strong diffusive latitudinal
 spreading of the toroidal flux underneath the activity belts can be
 inhibited by an inflow towards the toroidal field bands in the
 convection zone with a magnitude of several meters per second.}
{The inferred value of the turbulent magnetic diffusivity
  affecting the toroidal field agrees, to order of magnitude, with
  estimates based on mixing-length models for the solar convection
  zone. This is at variance with the requirement of flux-transport
  dynamo models. The inflows required to keep the toroidal field bands
  together before they approach the equator are similar to the inflows
  towards the activity belts observed with local helioseismology.}

\keywords{Magnetohydrodynamics (MHD) -- Sun: dynamo -- Sun: surface magnetism}
\authorrunning{Cameron \& Sch{\"u}ssler.}
\titlerunning{Turbulent diffusion of toroidal magnetic flux}
\maketitle

\section{Introduction}
\label{sec:intro}

Our understanding of the solar dynamo is based on theory,
helioseismology, and the observational clues that we can derive from the
emergence of the magnetic flux through the solar surface and its
subsequent evolution. A well studied theoretical model is the
Flux-Transport Dynamo (FTD), which explains the equatorward migration of
the activity belts in terms of advective transport of toroidal flux by
assuming a deep return flow of the observed near-surface poleward
meridional flow.  In order for this mechanism to work, the turbulent
magnetic diffusivity affecting the toroidal field needs to be low enough
to be overcome by the flow. Since simple estimates of the turbulent
magnetic diffusivity in the convection zone based on the mixing length
model give high values of the order of $1000$~km$^2$s$^{-1},$ most FTD
models therefore assume that the toroidal flux is generated mainly by
radial differential rotation associated with the tachocline and stored
in a region of much lower magnetic diffusivity at or below the bottom of
the convection zone \citep[for reviews, see,
e.g.][]{Dikpati:Gilman:2009,Charbonneau:2010, Charbonneau:2014,
Karak:etal:2014}. Low diffusivity, possibly together with downward
convective pumping \citep[e.g.][]{Jiang:etal:2013b, Lawson:etal:2015},
also prevents rapid loss of toroidal flux by radial diffusion
\citep{Zita:2010}. Alternatively, it has been suggested that the
required low values of the magnetic diffusivity affecting the toroidal
field could be provided within the convection zone through diffusivity
quenching by a sufficiently strong magnetic field
\citep[e.g.][]{Munoz-Jaramillo:etal:2011}. In any case, the toroidal
flux needs to be associated by a magnetic diffusivity that is orders of
magnitude lower than the estimates for the convection zone based on
mixing-length models.

In this paper, we use observations to infer the turbulent magnetic
diffusivity, $\eta_{\mathrm{t}}$, affecting the toroidal flux bands that
provide the source of the magnetic flux emerging to form sunspot groups.
We base our analysis on the properties of the time-latitude diagram of
sunspot emergence (butterfly diagram) derived from the RGO/SOON sunspot
record for solar cycles 12--23. We find that all cycles decline in the
same manner once the centers of their activity belts have approached the
solar equator to within their width at half maximum
(Section~\ref{sec:bf}). Interpreting this result in terms of toroidal
flux diffusing across the equator and cancelling with opposite-polarity
flux, we determine a range of values for $\eta_{\mathrm{t}}$ that is
within an order of magnitude with the mixing-length result for the
convection zone (Section~\ref{sec:Interpretation1}). This is in contrast
to the requirement of FTD models that the toroidal field should be
affected by a much lower magnetic diffusivity.

The high value of $\eta_{\mathrm{t}}$ would lead to severe spreading of the
activity belts before they have propagated into the vicinity of equator,
contrary to what is observed.  Such spreading could be prevented by a
latitudinal inflow towards the toroidal field bands. We estimate the
structure and amplitude of such a flow in Section~\ref{sec:inflow}.
It turns out that the properties of the inferred flow are similar to
those of the observed shallow inflows towards the activity belts
\citep{Gizon01,Gizon04,Zhao04}.

\section{Properties of the butterfly diagram}
\label{sec:bf}
\begin{figure*}
  \begin{center}
    \includegraphics[scale=0.65,clip = true]{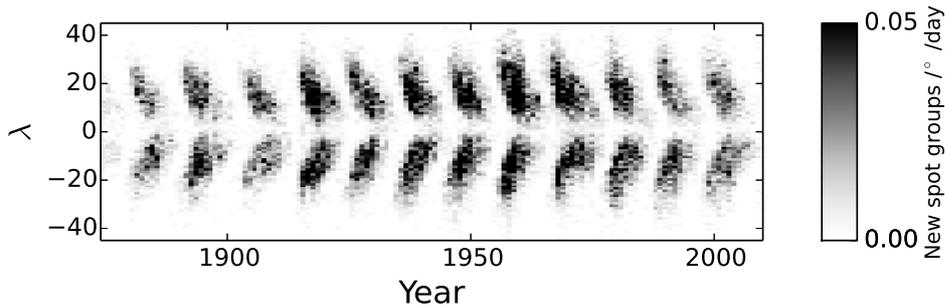}
    \caption{Time-latitude diagram (butterfly diagram) of the 
     rate of sunspot group emergence based on the RGO/SOON data sets.}
    \label{fig:bf}
  \end{center}
\end{figure*}

\begin{figure*}
  \begin{center}
\includegraphics[scale=0.35,clip = true]{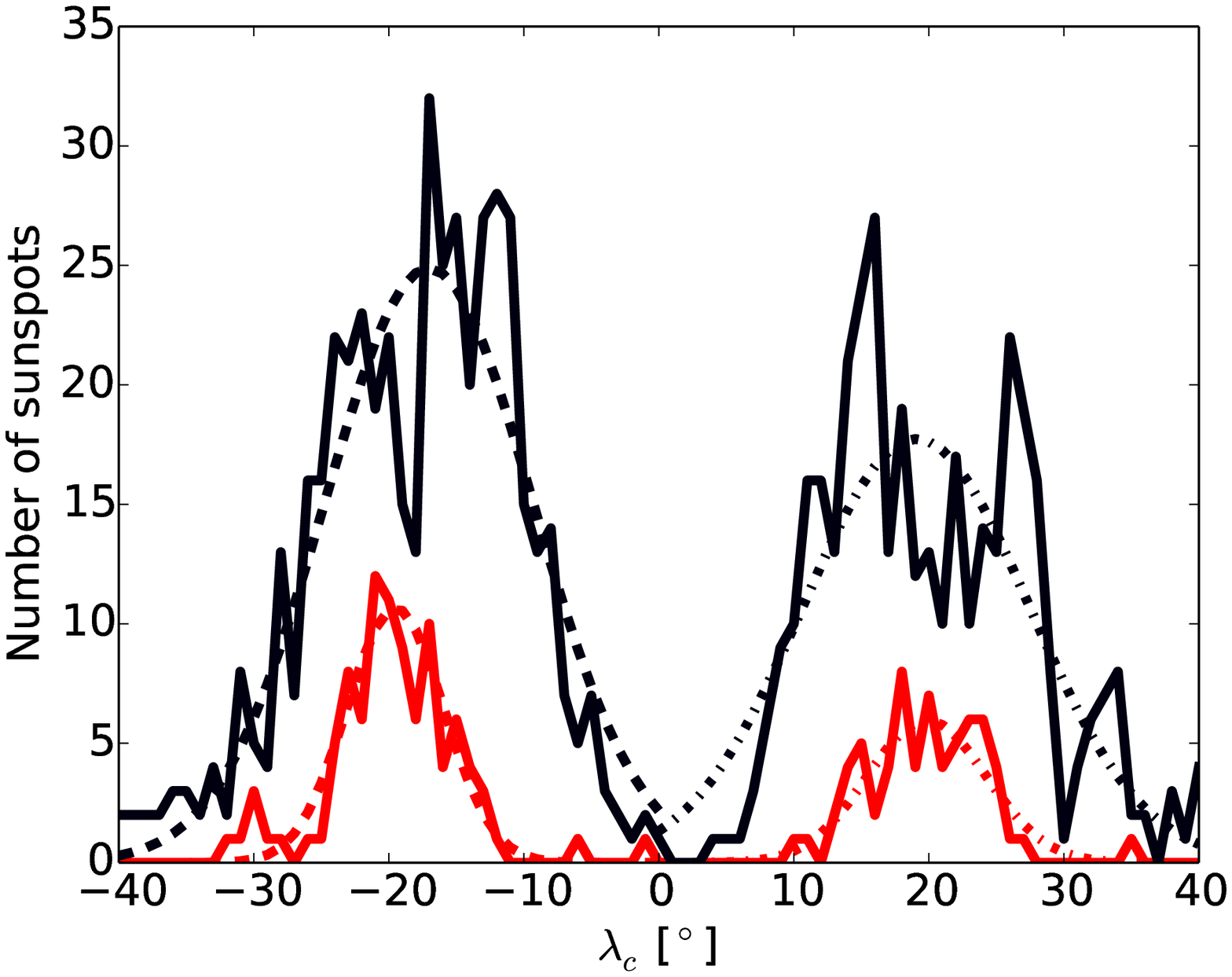}%
\includegraphics[scale=0.35,clip = true]{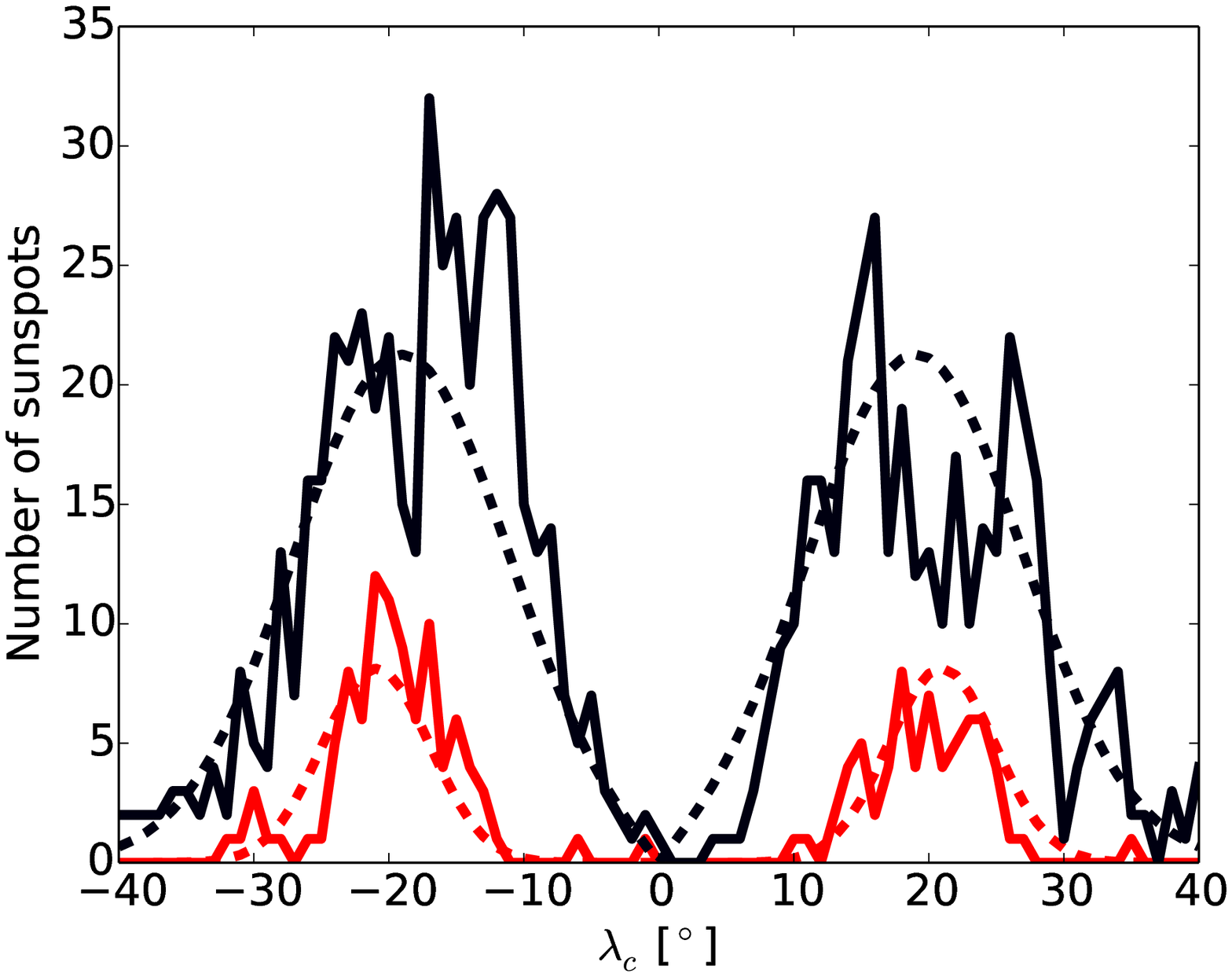}
    \caption{Latitudinal cuts through the butterfly diagram shown in
      Fig.~\ref{fig:bf} for the years 1903 (during the weak cycle 14,
      solid red curve) and 1957 (during the strong cycle 19, solid black
      curve). In the left panel, the dashed (dash-dotted) curve
      represents a Gaussian fit to the number of spot groups in the
      southern (northern) hemisphere. In the right panel, the two hemispheres
      were fitted together with a single Gaussian.}
    \label{fig:fit_all}
  \end{center}
\end{figure*}

Figure~\ref{fig:bf} shows the emergence rate of sunspot groups as a
function of time and latitude, based on the RGO/SOON sunspot group
records.  We constructed this `butterfly diagram' by binning the sunspot
groups in yearly and 1-degree bins, considering each group at the time
of its maximum area.
%
Similar to the analysis of \cite{2011ARep...55..911I}, we determined
yearly values of the central latitude and the full width at half
maximum of the butterfly wings, i.e., the migrating activity
belts.  We fitted Gaussians to the latitudinal profiles in both
hemispheres using the {\sl curve\_fit} routine of NumPy (version 1.8.1).
Examples of the original profiles and the Gaussian fits for the years
1903 (during the early phase of the weak cycle 17) and the year 1957
(during the early phase of the strong cycle 19) are shown in the left
panel of Figure~\ref{fig:fit_all}. We also fitted the emergence rates in
both hemispheres together (as a function of unsigned latitude); the
corresponding fit is shown in the right hand panel.  In most of the
following we will show the results for the fits based on the unsigned
latitude (i.e., considering both hemispheres together) because the noise
is lower.
{\added{The main deviation from the Gaussian shape occurs near the
    equator, where fewer sunspots are observed than indicated by the
    Gaussian fits. Such an asymmetry is expected if the two activity
    belts (or rather the underlying oppositely directed toroidal flux
    systems) interact destructively and cancel each other across the
    equator. }}

We determine three parameters describing the evolution of the activity
belt: amplitude $A(t)$, central latitude $\lambda_c(t)$, and its full
width at half maximum, $w(t)$. In addition, we consider the yearly
international sunspot number, $S(t)$ \citep[version 2.0][]{sidc}, all
for cycles 12 through 23. Because cycles overlap in time by a few years,
which is not accounted for in our Gaussian fitting, we follow
\cite{Solanki08} and consider the parameters only from years $2$ to $9$
during each cycle. Because we are interested in the relationship between
activity and the latitudinal distribution (described by $\lambda_c(t)$
and $w(t)$), we can choose to consider either $S(t)$ and $A(t)\times w(t)$
as measure of the activity level. We have tested
both choices and found the results to be similar. Here we present the
results based on $S(t)$ because as a measure of activity it has been
well studied and calibrated \citep{2014SSRv..186...35C}.  In the
following we therefore consider the relation between the three
quantities $\lambda_c(t)$, $w(t)$, and $S(t)$, which describe the evolution
of the activity belts during the cycles covered by the data.

\begin{figure}
  \begin{center}
    \includegraphics[scale=0.45,clip = true]{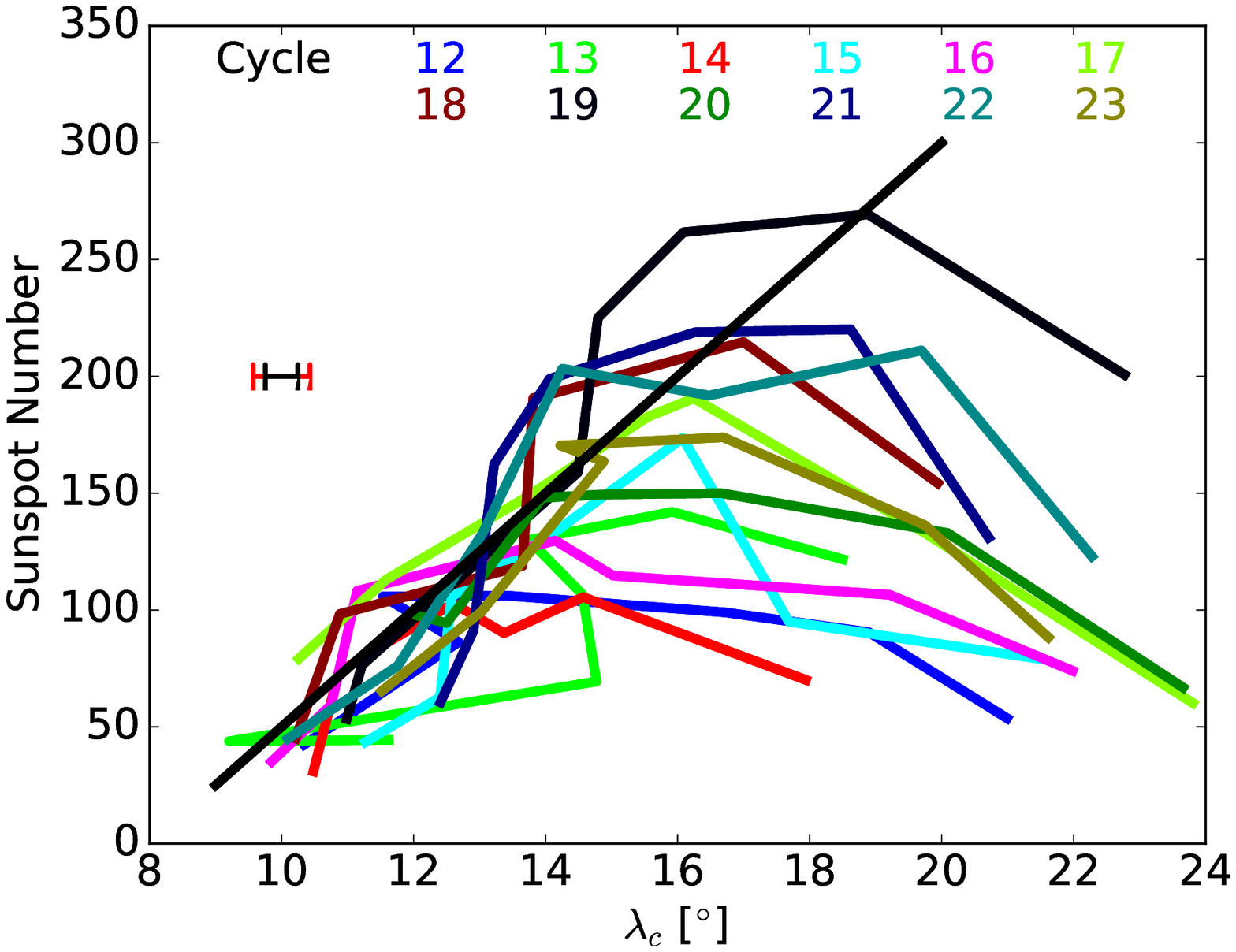}
    \caption{Yearly International Sunspot Number $S$ \citep{sidc} for
            cycles 12 to 23 as a function of the mean unsigned latitude
            of the sunspots each year. The line colors indicate the
            different cycles. The thin black curve is an approximate fit
            to the declining phase: $S=25 \lambda_c-200$.
            {\added{Error bars on the left side indicate standard
            deviations of $\lambda_c$ from the Gaussian fitting
            procedure carried out for each year: the maximum value is shown in red while the mean
            over all years is given in black.}} }
    \label{fig:lat_ssn}
  \end{center}
\end{figure}

Figure~\ref{fig:lat_ssn} shows the sunspot number, $S(t)$, as a function of
the central latitude of the activity belt, $\lambda_c(t)$, separately
for cycles 12 to 23.  This plot is similar to Figure~6d of
\citet{2014Ge&Ae..54..907I}, except that we use the central latitude
obtained from the fits for each year while they use an empirically
derived model of the central latitude.  This difference has no
substantial impact. We see that each cycle begins with a different
activity level at high latitude. As the activity belt propagates
towards the equator, the sunspot number at first increases and
then decreases sharply. The latitude at which the sunspot number begins
to decrease depends on the strength of the cycle, with stronger cycles
beginning to decline already when their activity belts are at higher
latitudes.  Because $\lambda_c$ moves towards the equator in the same
way for all cycles \citep{Waldmeier55, Hathaway11b}, the decrease at
high latitudes for stronger cycles is equivalent to a decrease earlier
in the cycle than for weaker cycles.  The fact that strong cycles begin
to decline earlier is known as the Waldmeier effect. After they begin to
decline, all the cycles show a similar sunspot number as a function of
their mean unsigned latitude.  The straight black line in
Figure~\ref{fig:lat_ssn} given by $S=25 \lambda_c-200$ gives a
rough approximation of this common declining phase.

\begin{figure}
  \begin{center}
    \includegraphics[scale=0.45,clip = true]{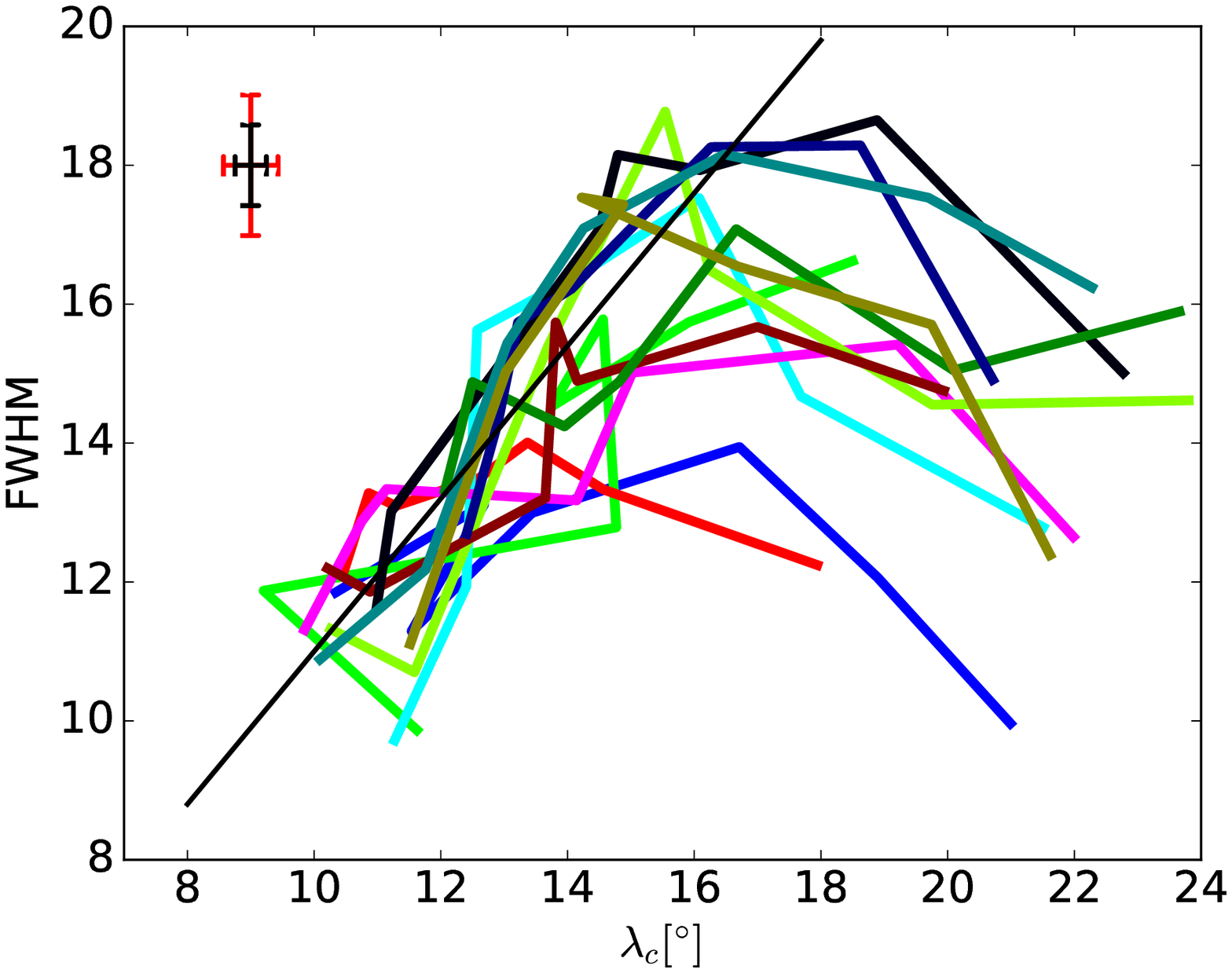}
    \caption{Full width at half maximum ($w$) of the activity belts as a
            function of the mean unsigned latitude of the sunspots each
            year. Different colors represent different cycles as in
            Fig.~\ref{fig:lat_ssn}. The thin black curve is an
            approximate fit to the declining phase:
            $\mbox{$w$}=1.1\lambda_c$.  {\added{Error bars on the top
            left indicate standard deviations of $\lambda_c$ and
            $w$, respectively, from the Gaussian fit procedure carried
            out for each year: maximum values are shown in red while the
            mean values over all years are given in black.}} }
    \label{fig:lat_hwhm}
  \end{center}
\end{figure}

Next we consider the relationship between the full width at half
maximum, $w(t)$, and the central latitude, $\lambda_c(t)$, of the
activity belts for cycles 12 to 23, which is shown in
Figure~\ref{fig:lat_hwhm}.  The belt width as a function of mean
unsigned latitude behaves in analogy to the sunspot number shown in the
previous figure: stronger cycles begin with a high $w$ that starts to
decrease at higher latitudes than during weaker cycles. In their declining
phases (at low latitudes), all cycles show a similar decrease of
$w$. The common behavior in the declining phase can be
approximated by the relation $w=1.1 \lambda_c$.

\begin{figure*}
  \begin{center}
    \includegraphics[scale=0.7,clip = true]{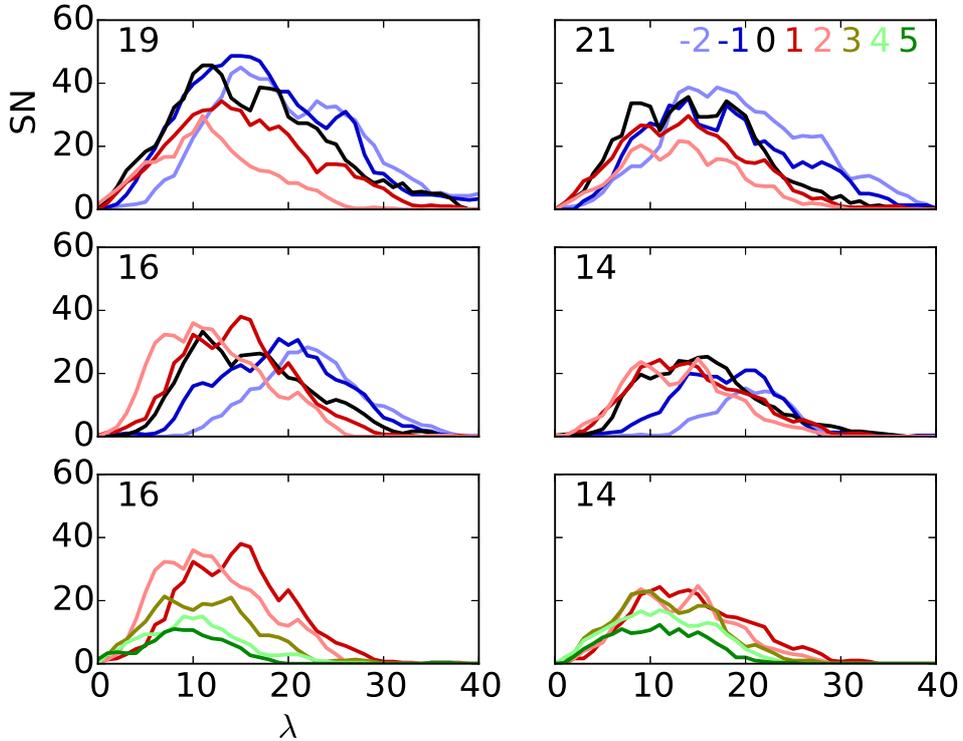}
    \caption{Latitudinal distribution of sunspots for several years
      during cycles 19 (upper left), 21 (upper right), 16 (middle and
      lower left) and 14 (middle and lower right). The black curves are
      for the years when the mean latitude of the sunspots was between
      $13^\circ$ and $15^\circ$. The colours of the other curves
      indicate the year relative to this reference year (see legend in
      the top left panel). The curves were smoothed with a 3 degree
      box-car filter. The late phases of cycles 14 and 16 are shown
      separately (bottom panels) in order to avoid too strong crowding
      of lines.}
    \label{fig:cycs_lat_years}
  \end{center}
\end{figure*}

An important point for the interpretation of the results is that the
decline of activity once the sunspot belts have approached the equator
to within their width affects all latitudes within the belts in the same
way; it is a universal decline, not just a decrease on its equatorward
side. This is illustrated in Fig.~\ref{fig:cycs_lat_years}, which shows
the evolution of the activity belts for the strongest cycles (numbers 19
and 21) and the weakest cycles (numbers 14 and 16) in the data
set. Relative to the year when $\lambda_c$ was between $13^\circ$ and
$15^\circ$ latitude (year zero, black lines), the plots show the
latitudinal distribution of sunspot activity in the two preceding and up
to five subsequent years. For the strong cycles, the decline begins at
year zero while for the weak cycles the belts continue to migrate
equatorwards for another two years with no significant change of
amplitude. The activity profiles for the following two years then show
an overall decrease: activity declines at all latitudes of the activity
belt while the shape stays roughly Gaussian.

At this point, we have seen that all cycles decline in a very
similar way: the activity belts propagate towards the equator at the
same rate, and both their amplitude and width can be approximated by
simple functions of their latitudinal distance from the equator. Cycles
grow at different rates, with stronger cycles beginning to decline
earlier (the Waldmeier effect). But once they begin to decline,
cycles all decline in a similar way and the activity decreases
over the whole width of the activity belts.  We discuss
the implications of these results in the next section.

\section{Interpretation and estimate of $\eta_{\mathrm{t}}$}
\label{sec:Interpretation1}

The basic result obtained in the previous section is that all cycles
decline in the same way.
Figs.~\ref{fig:lat_ssn}--\ref{fig:cycs_lat_years} indicate that the
activity belts propagate towards the equator showing increasing or
approximately constant levels of activity until their distance from the
equator becomes about equal to their width. Thereafter the activity
strongly decreases across the whole latitude range covered by the belts.
Stronger cycles begin with a larger width of their activity belts and
thus start to decline earlier than weaker cycles. Once they begin to
decline, all cycles follow the same pattern in terms of amplitude and
latitudinal distribution. These properties suggest an interpretation of
the declining phase in terms of cross-equator diffusion and cancellation
of toroidal magnetic flux. We consider this possibility by a
quantitative analysis, which provides an estimate of the magnitude of
the relevant magnetic diffusivity.

For our analysis, we use radians as unit for $\lambda$ and $w(t)$.  We
consider the distribution of sunspots as a function of latitude and
time, $N(\lambda,t)$, assuming that the activity belts in both
hemispheres are symmetric with respect to the equator.  We model this
quantity as a Gaussian with full width at half maximum given by $w(t)$,
centered at $\lambda_c$ and with a total area given by the sunspot
number $S/2$, viz.
\begin{eqnarray}
N(\lambda,t) = \frac{S(t)}{2 \sigma \sqrt{2 \pi}}   
 \exp{\left(-\frac{(\lambda-\lambda_c)^2}{ 2 \sigma^2}\right)}
\label{eqn:N} 
\end{eqnarray}
where $\sigma(t)=w(t) / \sqrt{8 \ln 2}$.

We assume that the latitudinal profiles of the activity belts represent
those of the underlying bands of toroidal magnetic flux from which the
bipolar magnetic regions and sunspots emerge. This appears to be a
reasonable assumption since the activity belts do not overlap and even at low
latitudes it is uncommon to see sunspot groups not obeying Hale's
polarity rules. If we assume further that the emerging flux (represented
by the sunspot number) is proportional to the radially integrated
toroidal flux density at each latitude, we obtain
\begin{eqnarray}
  \int_{0.7 R_{\odot}}^{R_{\odot}} r B_\phi(r,\lambda,t) \mathrm{d}r 
   = \alpha N(\lambda,t),
\label{eqn:spots_to_bphi}
\end{eqnarray}
where $B_{\phi}(r,\lambda,t)$ is the longitudinally averaged azimuthal
component of the magnetic field, $\alpha$ is a constant of
proportionality (which drops out of the final result), and $R_{\odot}$
is the solar radius.

\begin{figure}
\begin{center}
\includegraphics[scale=0.42,clip = true]{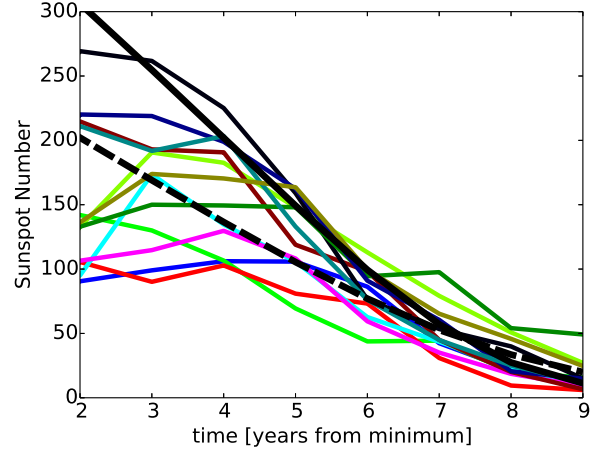}
\caption{Sunspot Number as a function of time of minimum activity.
       Different colors represent different cycles as in
       Fig.~\ref{fig:lat_ssn}.}
\label{fig:model}
\end{center}
\end{figure}

We interpret the observational results illustrated in
Figs.~\ref{fig:lat_ssn}--\ref{fig:cycs_lat_years} in terms of diffusive
cancellation of the oppositely oriented toroidal flux bands across the
equator. 
{\added{This interpretation is motivated by the above result that the
    decline in activity each cycle begins only when the activity belts
    come within their width of the equator, but from then on affects the
    whole activity belt. This behavior suggests that the loss of
    toroidal flux across the equator dominates over the losses due to
    cancellation at the axis of rotation as well as due to downward pumping
    of flux toward stable layers near the bottom of the convection zone.}}
We consider the rate, $D_{\mathrm{eq}}$, at which
toroidal flux in the northern hemisphere is transported across the
equatorial plane by diffusion. Applying Stokes' theorem to the
integrated toroidal flux over a hemisphere yields
\begin{eqnarray}
D_{\mathrm{eq}}=\int_{0.7 R_{\odot}}^{R_{\odot}}\left.
  \frac{\eta_{\mathrm{t}}(r)}{r} 
  \frac{\partial B_\phi(r,\lambda,t)}
  {\partial \lambda}\right\vert_{\lambda=0} dr,
\label{eqn:full}
\end{eqnarray}
where $\eta_{\mathrm{t}}(r)$ is the turbulent magnetic diffusivity.
Since the depth-dependence of the toroidal field (i.e., the storage
location of the toroidal flux) is unknown, we rewrite
Eq.~(\ref{eqn:full}) in the form
\begin{eqnarray}
D_{\mathrm{eq}}=\left(\frac{\eta}{R^2}\right)_{\mathrm{e}}
   \frac{\partial}{\partial\lambda} \left( \left. 
   \int_{0.7 R_{\odot}}^{R_{\odot}} 
   r B_\phi(r,\lambda,t) dr \right) \right\vert_{\lambda=0}
\end{eqnarray}
where $(\eta/R^2)_{\mathrm{e}}^{-1}$ is an effective diffusion time of
the toroidal flux. If the flux is stored in a narrow radial layer around
$r=R_0$, then $(\eta/R^2)_{\mathrm{e}}=\eta_{\mathrm{t}}(R_0)/R_0^2$ and the
relevant turbulent magnetic diffusivity affecting the toroidal field is
given by $\eta_{\mathrm{t}}(R_0)$. For more general distributions,
$(\eta/R^2)_{\mathrm{e}}$ represents the effective quantity describing
the diffusion of the toroidal flux.  Using Eq.~(\ref{eqn:spots_to_bphi}),
we obtain
\begin{eqnarray}
D_{\mathrm{eq}}=\left(\frac{\eta}{R^2}\right)_{\mathrm{e}}
   \alpha\left. \frac{\partial  N}{\partial \lambda} \right\vert_{\lambda=0}
\end{eqnarray}
and substituting Eq.~(\ref{eqn:N}) we find
\begin{eqnarray}
D_{\mathrm{eq}}=\left(\frac{\eta}{R^2}\right)_{\mathrm{e}} \alpha
   \frac{\lambda_c S}{2 \sigma^3 \sqrt{2 \pi}}  
   \exp{\frac{-\lambda_c^2}{2 \sigma^2}}. 
\label{eqn:rate}
\end{eqnarray}
for the rate at which the northern hemisphere loses toroidal flux by
turbulent diffusion over the equator. Since opposite-polarity flux from
the southern hemisphere diffuses into the northern hemisphere at the
same rate, $D_{\mathrm{eq}}$ is actually half the rate at which the
toroidal flux in the northern hemisphere decreases. Therefore we have
\begin{eqnarray}
  D_{\mathrm{eq}}
  &=&\frac{1}{2}\frac{d}{dt} \int_0^{\pi/2} \int_{0.7 R_{\odot}}^{R_{\odot}} 
      r B_\phi(r,\lambda,t) \mathrm{d}r d\lambda\nonumber\\
  &=&\frac{\alpha}{2} \frac{d}{dt} \int_0^{\pi/2} N(\lambda,t)d 
     \lambda \nonumber\\
  &=&\frac{\alpha}{4}\frac{d S}{d t}.
\label{eqn:rate_final}
\end{eqnarray}
Equating Eqs.~(\ref{eqn:rate}) and (\ref{eqn:rate_final}) yields
\begin{eqnarray}
\frac{d S}{d t}=  4 \left(\frac{\eta}{R^2}\right)_{\mathrm{e}}
                  \frac{\lambda_c S}{2 \sigma^3 \sqrt{2 \pi}} 
                  \exp{\frac{-\lambda_c^2}{2 \sigma^2}}.
\label{eqn:ODE}
\end{eqnarray}
To model the declining phase, we use the empirical relations
obtained in Sec.~\ref{sec:bf}, namely
\begin{eqnarray}
  \lambda_c &=&  (S+200)/25\times \pi/180 \nonumber\\
            &=& 7\times10^{-4}(S+200)
\end{eqnarray}
from Figure~\ref{fig:lat_ssn} and 
\begin{eqnarray}
  \sigma&=& w/\sqrt{8 \ln{2}} =0.47\lambda_c \nonumber\\
  &=& 3.3 \times10^{-4}(S+200)
\end{eqnarray}
from Figure~\ref{fig:lat_hwhm}.  Equation~(\ref{eqn:ODE}) then is a
nonlinear ODE, which describes the evolution of the sunspot number
during the declining phase of a cycle. We solve this ODE numerically.
Figure~\ref{fig:model} shows the observed decrease in $S$ as a
function of time (counted from activity minimum) for each cycle.  The
thick black curves show estimates for the evolution of $S(t)$ based on
Eq.~(\ref{eqn:ODE}) corresponding to two values of
$(\eta/R^2)_{\mathrm{e}}R_\odot^2$: $450$~km$^{2}$s$^{-1}$ (solid black curve)
and $300$~km$^{2}$s$^{-1}$ (dashed black curve). Since $1 \le
{R_\odot^2}/{R^2} \la 2$, we obtain values in the range of
$150$~km$^{2}$s$^{-1}$ to $450$~km$^{2}$s$^{-1}$ for the effective
turbulent diffusivity affecting the toroidal flux bands, with most of
the uncertainty being due to the uncertainty in the depth at which the
toroidal flux is located.

One may imagine alternative explanations of the observed evolution
of the activity belts discussed in Sec.~\ref{sec:bf}, which would allow
the toroidal field to be stored in a low-diffusivity environment
somewhere below the convection zone proper. One possibility would be
magnetoshear instabilities in the tachocline
\citep[e.g.][]{Cally:etal:2003, Gilman:etal:2007}. However, these
instabilities are not restricted to low latitudes and also would require
a very peculiar nonlinear development in order to reproduce the uniform
decline of activity throughout the whole activity belt and its onset
depending on the width of the belt. The same problem is faced when
assuming a field-dependent low-latitude upflow part of the meridional
circulation: it needs to be of a very contrived field-dependent form to
be able to reproduce the observed evolution of the activity belts, i.e.,
mimick a diffusion process.

\section{Inflows maintaining the activity belts}
\label{sec:inflow}

The high value for the turbulent magnetic diffusivity obtained here
raises a problem.
    {\added{Our estimated range of diffusivities,
    150--450~km$^2$s$^{-1}$, corresponds to
    5.7$^2$--9.8$^2$~deg$^2$year$^{-1}$.  If diffusion were the only
    process operating, then even with the lower diffusivity an initial
    Gaussian distribution with a FWHM of $14^\circ$ degrees would
    increase its width by about $9.4^{\circ}$ after only one year. This
    is inconsistent with the observed development of the activity belts
    during the rising phase of a cycle (cf. Fig.~\ref{fig:lat_hwhm}),
    which show an increase of their width by only roughly 2--3$^{\circ}$
    per year.  To restrict the spread of the wings to $3^{\circ}$ or
    less would require the diffusivity to be less than
    40~km$^2$/s$^{-1}$. Higher values of the diffusivity therefore
    require the existence of an antidiffusive effect to counter too
    strong spreading of the activity belts. One possibility is to
    suppose that the toroidal field is affected by a latitudinal flow
    (inflow) towards the activity belt such that diffusive spreading is
    (partially) balanced by advection.  Here we consider the simple
    kinematic problem with the aim of estimating the properties of an
    inflow that would balance the diffusive spreading.}}

To estimate the geometry and amplitude of such a flow field,
we consider the induction equation for the azimuthally averaged
magnetic field, ${\bf B}$, 
\begin{eqnarray}
\frac{\partial {\bf B}}{\partial t}= 
      \nabla \times ( {\bf U} \times  {\bf B}) - 
      \nabla \times \eta_{\mathrm{t}}\nabla \times  {\bf B}.
\end{eqnarray}
The rate at which the activity belts propagate towards the equator is
about 1--2 degrees per year, while the expected diffusive spreading is
substantially higher. Therefore we can make the approximation that the
diffusion is balanced by the advective transport, i.e.
\begin{eqnarray}
  \nabla \times ( {\bf U} \times  {\bf B}) = 
  \nabla \times \eta_{\mathrm{t}}\nabla \times  {\bf B}.
\end{eqnarray}
Integration yields
\begin{eqnarray}
 {\bf U} \times  {\bf B} = \eta_{\mathrm{t}}\nabla \times  {\bf B},
\label{Eq:integrated}
\end{eqnarray}
where we have set the arbitrary scalar potential to zero.  The radial
component of Eq.~(\ref{Eq:integrated}) reads
\begin{eqnarray}
   U_{\lambda} B_{\phi}=
   \frac{\eta_{\mathrm{t}}}{r \sin(\lambda)}
   \frac{\partial \sin(\lambda)B_{\phi}}{\partial \lambda}\,.
\label{Eq:U}
\end{eqnarray}
On the left-hand side of this equation, we have neglected the term
$-U_{\phi} B_{\lambda}$, which describes the generation of toroidal
field by differential rotation. This is justified since we show in
Appendix~\ref{app} that the rate of flux change due to differential
rotation in the low latitudes covers at most 10\% of the required value
derived from the observed butterfly diagram of sunspot
emergence. 

Similar to the determination of the effective diffusion time,
$(\eta/R^2)_{\mathrm{e}}^{-1}$, in Sec.~\ref{sec:Interpretation1} we
derive an effective advection time, $(U_\lambda/R)_{\mathrm{e}}^{-1}$,
of the toroidal flux owing to latitudinal inflows. Integrating
Eq.~(\ref{Eq:U}) over radius and introducing the effective time scales
we obtain
\begin{eqnarray}
    (U_\lambda/R)_{\mathrm{e}} \int_{0.7 R_{\odot}}^{R_{\odot}} 
  &r& B_\phi(r,\lambda,t) \mathrm{d}r = \nonumber\\
  &=&
     \frac{(\eta/R^2)_{\mathrm{e}}}{\sin\lambda}
     \frac{\partial}{\partial\lambda} \left( \sin\lambda
     \int_{0.7 R_{\odot}}^{R_{\odot}} 
     r B_\phi(r,\lambda,t) \mathrm{d}r \right) \,.
\label{Eq:U2}
\end{eqnarray}
Using Eq.~(\ref{eqn:spots_to_bphi}) then yields
\begin{equation}
    (U_\lambda/R)_{\mathrm{e}} = 
    \frac{(\eta/R^2)_{\mathrm{e}}}{N(\lambda,t)\sin\lambda}
    \frac{\partial}{\partial\lambda}\left[ N(\lambda,t)\sin\lambda)
    \right] \,,
\label{Eq:U3}
\end{equation}
connecting the effective inflow with the time-latitude development of
the sunspot emergence rate $N(\lambda,t)$ shown in Figure~\ref{fig:bf}.
If the toroidal flux is stored in a narrow radial layer around $r=R_0$,
then $(U_\lambda/R)_{\mathrm{e}}=U_\lambda(R_0)/R_0$ and the relevant
inflow velocity is given by $U_\lambda(R_0)$.

\begin{figure}
  \begin{center}
    \includegraphics[scale=0.45,clip = true]{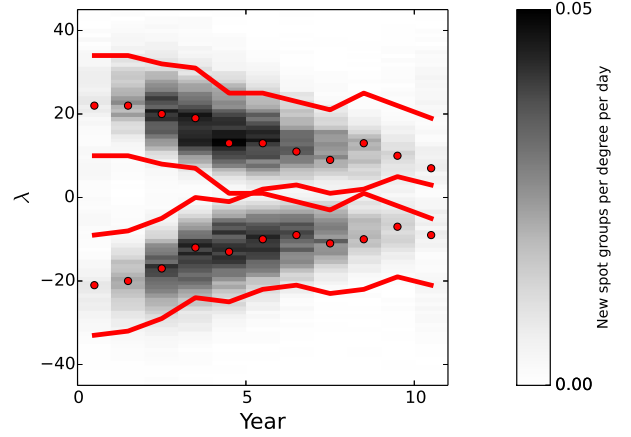}
    \caption{The average activity belts, based on aligning the
    butterfly wings shown in Figure~\ref{fig:bf}. The alignment was
    based on the year in which the average latitude of the wings is
    closest to 15 degrees. The red dots show the latitudes at which the
    activity belts are centered each year, and the red lines show the
    central latitude $\pm 12^{\circ}$.}
    \label{fig:bf_average}
  \end{center}
\end{figure}

\begin{figure}
  \begin{center}
    \includegraphics[scale=0.45,clip = true]{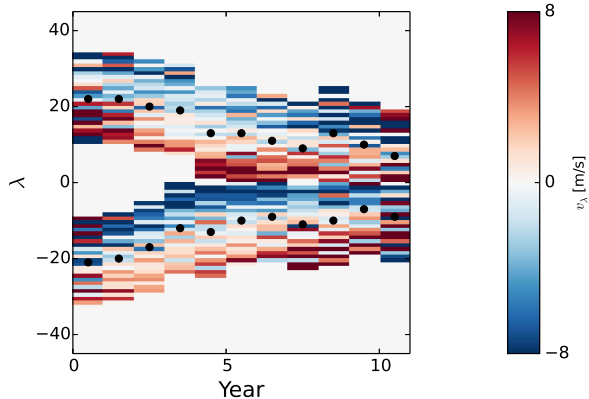}
    \caption{The velocity required to prevent the activity belts from
     diffusing in latitude. The velocity is only shown in the latitudes
     within the red lines shown in Figure~\ref{fig:bf_average}.}
    \label{fig:vel}
  \end{center}
\end{figure}

We evaluated Eq.~(\ref{Eq:U3}) on the basis of an avererage butterfly
diagram in order to reduce the noise. All cycles in our dataset were
aligned in time with respect to the year during which the mean latitude
of the sunspot groups in each wing is closest to 15$^{\circ}$. The
choice of 15$^{\circ}$ is somewhat arbitrary, but all cycles showed a
reasonable number of sunspots near this latitude and the results are not
strongly sensitive to this choice.  The average butterfly diagram
constructed in this way is shown in Figure~\ref{fig:bf_average}.  We
used the data for this average cycle and Eq.~(\ref{Eq:U3}) together with
Eq.~(\ref{eqn:spots_to_bphi}) to obtain
$(U_\lambda/R)_{\mathrm{e}}R_\odot$ for an intermediate value of the
magnetic diffusivity given by $(\eta/R^2)_{\mathrm{e}}R_\odot^2 =
300\,$km$^2$s$^{-1}$. The result is shown in Figure~\ref{fig:vel}.
Considering the ranges for
$(\eta/R^2)_{\mathrm{e}}R_\odot^2=150\dots450\,$km$^2$s$^{-1}$ and
$R/R_\odot=0.7\dots 1$, the corresponding amplitudes of $U_\lambda$
become between 65\% lower ($\simeq 3\,$m$\,$s$^{-1}$) and 50\% higher
($\simeq 12\,$m$\,$s$^{-1}$).

Note that we estimate $U_{\lambda}$ under the assumption that advection
exactly balances the outward diffusion. This is no longer valid near the
equator in the declining phase of a cycle, when the activity belts
approach each other. Since the inflows probably extend  beyond
the activity belts, the overlapping between the flow systems of the two
hemispheres reduces the inflows and thus allows for cross-equatorial
diffusion of toroidal flux.  The high velocities near the equator seen
in Figure~\ref{fig:vel} should therefore not be considered as realistic.

The inflows derived here are similar in both extent and amplitude to the
actually observed shallow inflows into the active region latitudes
\citep{Gizon01,Gizon04,Zhao04}. Furthermore, \citet{Komm15} showed that
the inflows are part of the extended solar cycle, beginning at mid
latitudes (about 40 degrees). The inflows then propagate equatorward
together with the activity belts. The extension in depth of these
flows is unclear at the moment. \citet{Liang:Chou:2015} report that the
helioseismic signature of the inflows is present down to
0.88--0.93~$R_\odot$, and partly present at depths from
0.75--0.88~$R_\odot$. On the other hand, \citet{Gizon:Rempel:2008}
report a return flow at a depth of about 60 Mm ($\simeq 0.91 R_\odot$).
Such rather shallow inflows could result from cooling associated with
the excess brightness of facular areas \citep{Spruit03}. Deep inflows
could be driven by thermal perturbations caused by toroidal field
amplification near the base of the convection zone
\citep{Rempel:2003}. In fact, there are helioseismic indications of
cycle-related changes of the sound speed in this region
\citep{Baldner:Basu:2008}.

Since the inflows probably extend only over a limited depth range in
the convection zone, we expect that part of the toroidal flux is spread
over the convection zone by radial turbulent diffusion and an outward
return flow.  The effect of a deep return flow (in case of an inflow
confined to the upper convection zone) is negligible since, owing to
mass conservation and the strong increase of density with depth, its
speed is much lower than that of the inflow. Radial turbulent diffusion
cannot be neglected in this case, but the solution of a simple 2D
advection-diffusion problem shows that still a substantial part of the
toroidal flux is kept confined in the convergence zone of the inflows.

\section{Conclusion}

Solar cycles begin to decline when the distance of the center of the
activity belts from the equator becomes about equal to their width (at
half maximum). From this time onward, activity decreases across the
whole latitude range covered by the belts and all cycles decline in the
same way. Stronger cycles show wider activity belts and thus start to
decline earlier than weaker cycles. This is the essence of the
Waldmeier effect \citep{Waldmeier55}.

We interpret this result in terms of oppositely directed toroidal flux
bands in each hemisphere that diffuse and cancel across the equator.
The observational results are consistent with a diffusion process
with effective value of the turbulent diffusivity acting on the toroidal
magnetic field in the range $150$~km$^{2}$s$^{-1}$ to
$450$~km$^{2}$s$^{-1}$. This is consistent (to order of magnitude) with
the mixing length theory estimate, but inconsistent with the much lower
values required by flux transport dynamos. Other interpretations of the
observations in terms of magnetoshear instability or magnetic-field
dependent upflows would require rather contrived properties of these
processes, for which there is no evidence.

The flow field required to maintain the butterfly wings against the
effects of diffusion is an inflow of a few meters per second, centered
on the activity belts. Its spatial extent is at least $12^{\circ}$ on
both sides of the activity belt.  Inflows with similar properties
have been observed in the upper part of the convection zone, but may
extend also to deeper layers.

\section*{Acknowledgments}
         {We acknowledge fruitful discussions with Matthias Rempel, Mark
           Cheung, Jie Jiang, and Emre I{\c s}ik.  NSO/Kitt Peak data
           used here are produced cooperatively by NSF/NOAO, NASA/GSFC,
           and NOAA/SEL. This work utilizes SOLIS data obtained by the
           NSO Integrated Synoptic Program (NISP), managed by the
           National Solar Observatory, which is operated by the
           Association of Universities for Research in Astronomy (AURA),
           Inc. under a cooperative agreement with the National Science
           Foundation. The synoptic magnetograms were downloaded from
           the NSO digital library, http://diglib.nso.edu/ftp.html. The
           sunspot numbers were obtained from WDC-SILSO, Royal
           Observatory of Belgium, Brussels, http://sidc.oma.be/silso.
           The RGO and SOON datasets were downloaded from
           http://solarscience.msfc.nasa.gov/greenwch.shtml. 

\bibliographystyle{aa}
\bibliography{ISSI_SFT}

\begin{appendix}
\section{Generation of toroidal flux by differential rotation}
\label{app}
In order to estimate the effect of low-latitude generation of toroidal
flux by differential rotation in the declining phase of an activity
cycle, we consider the time derivative of the axisymmetric toroidal
magnetic flux, $\int_{R_{T}}^{R_{\odot}} B_{\phi} r \mathrm{d}r$,
radially integrated from the top of the tachocline, $R_T$, to the
surface, $R_{\odot}$, as a function of latitude. This puts aside the
question where the toroidal flux is located in radius, but still
requires us to make some assumptions on the internal structure of the
magnetic field. Firstly, we assume that turbulent/convective pumping
expels the horizontal components of the field from the near-surface
shear layer (NSSL), so that the field there becomes purely
radial. Secondly, we assume that the radial shear is zero between the
NSSL and the top of the tachocline, and, thirdly, we assume that the
magnetic field does not penetrate into the tachocline. Owing to the
prolate shape of the tachocline \citep[e.g.][]{Antia:Basu:2011}, the last
assumption is somewhat questionable in higher latitudes, but here we are
mainly interested in the flux generation in the low-latitude activity
belts,
{\added{ where the main part of the tachocline is located below the
         convection zone. In addition, toroidal flux generation in the
         tachocline faces the problem of maintaining the required radial
         shear in the presence of a substantial Lorentz force and weak
         (or even absent) convection \citep{Gilman:Rempel:2005,
         Spruit11}}}.

In what follows we use a proper right-handed system of spherical
coordinates, $(r,\theta,\phi)$. The contribution of differential
rotation, $U_{\phi}=\Omega(r,\theta)\,r\sin\theta$, to the evolution of
the azimuthally averaged toroidal field, $B_{\phi}$, is given by
\begin{eqnarray}
  \frac{\partial B_\phi}{\partial t}&=&\frac{1}{r} 
  \left[\frac{\partial (r U_{\phi} B_r)}{\partial r}
       +\frac{\partial (U_{\phi} B_{\theta})}{\partial\theta}\right]\nonumber \\
  &=&   \frac{1}{r} \left[ \frac{\partial}{\partial r}
      ( \Omega\, r^2 \sin\theta\, B_r) + \frac{\partial}{\partial\theta}
      ( \Omega\, r \sin\theta\, B_\theta)\right]\,.  
\end{eqnarray}
Integrating over radius yields 
\begin{eqnarray}
 \frac{\partial}{\partial t}\int_{R_T}^{R_{\odot}}B_{\phi} r\, \mathrm{d}r
  =&\int_{R_T}^{R_{\odot}}& \sin\theta\, 
  \frac{\partial (\Omega r^2 B_r)}{\partial r} \,\mathrm{d}r \nonumber \\
  &+&
  \frac{\partial}{\partial \theta} \int_{R_T}^{R_{\odot}}
  \Omega \sin\theta\,  B_{\theta}r  \,\mathrm{d}r 
\end{eqnarray}
Since we assume $B_r=0$ at $r=R_T$, we have
\begin{equation}
  \int_{R_T}^{R_{\odot}}  \sin\theta \, 
  \frac{\partial (\Omega r^2 B_r)}{\partial r} \mathrm{d}r
   = \left. \left[\sin\theta\, \Omega r^2 B_r \right] 
   \right\vert_{R_\odot} \,.
\end{equation}
With our assumptions that $B_\theta=0$ in the NSSL and that $\Omega$
is independent of $r$ between $R_T$ and $R_{\rm NSSL}$, we obtain
\begin{eqnarray}
  \frac{\partial}{\partial t}\int_{R_T}^{R_{\odot}}B_{\phi} r\, \mathrm{d}r
  = & &\left. \left[\sin\theta\, \Omega r^2 B_r \right]\right\vert_{R_\odot}
  \nonumber \\
  &+&
  \frac{\partial }{\partial \theta} \left(\left.\Omega\right\vert_{\rm NSSL} 
  \int_{R_T}^{R_{\rm NSSL}} \sin\theta\,  B_{\theta}r \mathrm{d}r \right) \,,
\label{app_eq_4}
\end{eqnarray}
where $R_{\rm NSSL}$ indicates the bottom of the NSSL and we write
$\Omega\vert_{\rm NSSL}=\Omega(R_{\rm NSSL},\theta)$.  The condition
$\nabla \cdot {\bf{B}}=0$ for the azimuthally averaged field implies
\begin{eqnarray}
r \frac{\partial}{\partial \theta} (B_{\theta} \sin\theta)=
-\sin\theta \frac{\partial}{\partial r}(r^2B_r)\,.
\end{eqnarray}
Integration of this equation over radius yields
\begin{eqnarray}
\frac{\partial}{\partial \theta} \int_{R_T}^{R_{\rm NSSL}} B_{\theta} 
   \sin\theta\, r \mathrm{d}r&=&
 -\sin\theta \int_{R_T}^{R_\odot}
  \frac{\partial(r^2B_r)}{\partial r}\mathrm{d}r \nonumber \\
  &=&\left. -\sin\theta\, R_\odot^2 B_r\right\vert_{R_\odot}\,.
\end{eqnarray}
where we have used $B_\theta=0$ in the shear layer.
By integrating over $\theta$ we obtain
\begin{eqnarray}
\int_{R_T}^{R_{\rm NSSL}} B_{\theta} \sin\theta\, r \mathrm{d}r  
 &=&-\int_0^{\theta}\left.\sin\theta \,
    R_{\odot}^2 B_r\right\vert_{R_{\odot}}\mathrm{d}\theta.
\end{eqnarray}
We now substitute this relation into the last term of
Eq.~(\ref{app_eq_4}), using the assumptions that $\Omega$ is independent
of $r$ below the NSSL and $B_{\theta}=0$ within the NSSL, to obtain
\begin{eqnarray}
\frac{\partial}{\partial t} \int_{R_T}^{R_{\odot}}B_{\phi} r \mathrm{d}r
  &=& \left.\left(\sin\theta\, \Omega r^2 B_r \right)\right\vert_{R_\odot}
    -  \frac{\partial}{\partial \theta}   \Omega\vert_{R_{\rm NSSL}} 
  \int_0^{\theta}\sin\theta\, R_{\odot}^2 B_r\vert_{R_{\odot}}
  \mathrm{d}\theta  \nonumber \\
  &=& \left.\left(\sin\theta\, \Omega r^2 B_r \right)\right\vert_{R_\odot}
  - \Omega\vert_{R_{\rm NSSL}} \left.\left(\sin\theta \,
   R_{\odot}^2B_r\right)\right\vert_{R_{\odot}} \nonumber \\ 
  & &- \left(\frac{\partial \Omega\vert_{R_{\rm NSSL}}}
  {\partial \theta}\right)
  \int_0^{\theta} \sin\theta\, R_{\odot}^2 B_r\vert_{R_{\odot}}
  \mathrm{d}\theta \,.
\end{eqnarray}
Finally we obtain
\begin{eqnarray}
  \frac{\partial \int_{R_T}^{R_{\odot}}B_{\phi} r \mathrm{d}r}{\partial t}  
  =& & \sin\theta\,  R_\odot^2 B_r\vert_{R_{\odot}} 
  \left( \Omega\vert_{R_\odot}- \Omega\vert_{R_{\rm NSSL}} \right)\nonumber \\ 
  &-& \frac{\partial \Omega\vert_{R_{\rm NSSL}}}{\partial \theta}
    \int_0^{\theta} \sin\theta\, R_{\odot}^2 B_r\vert_{R_{\odot}}
  \mathrm{d}\theta \,.
\end{eqnarray}
The first term on the right-hand side of this equation corresponds to
toroidal flux generation by radial differential rotation in the NSSL,
while the second term describes the generation by latitudinal
differential rotation. We know the value of both $\Omega\vert_{R_{\rm
NSSL}}$ and $\Omega\vert_{R_{\odot}}$ from observations. For the surface
rotation rate we use the synodic rate for magnetic fields of
\cite{Hathaway11},
\begin{eqnarray}
\label{eqn:omega_rsun}
  \Omega\vert_{R_{\odot}}&=& 14.30-1.98 \cos^2\theta-2.14 
  \cos^4\theta \mbox{\hspace{0.2cm}} [ ^{\circ}/\mbox{day} ]\,.
\end{eqnarray}
For the rotation rate at the base of the NSSL we we add a
latitude-independent value of $0.53^{\circ}/$day to the near-surface helioseismic
result of \cite{Schou_etal98}.
This is motivated by the results of \citet{Barekat14}, indicating that
the strength of the NSSL is independent of latitude.

The contributions to the generation of toroidal flux from both the
radial and latitudinal shear can then be evaluated using azimuthally
averaged KPNO/VTT and SOLIS synoptic maps and are shown in
Figure~\ref{fig:KPNO} (upper panels). The lower panels of this figure
compare the combined flux generation by radial and latitudinal shear
with the change of integrated toroidal flux as estimated from the
sunspot emergence rate according to Eq.~(\ref{eqn:spots_to_bphi}) in the
main part of the paper.  Here we chose the constant of proportionality,
$\alpha$, such that the result matches the total toroidal flux at
activity maximum ($\simeq 5\cdot10^{23}\,$Mx/hemisphere) as estimated according to
\citet{Cameron15} after integration in latitude.  Considering the
different scale ranges of the lower panels in Figure~\ref{fig:KPNO}, it
is clear that low-latitude generation of toroidal flux by differential
rotation falls short of the required flux changes in the activity belts
by at least a factor of 10. Incidentally, this result also indicates
that the latitudinal propagation of the sunspot zone cannot be ascribed
to a dynamo wave along the NSSL.

\begin{figure*}\includegraphics[scale=0.33,clip = true]{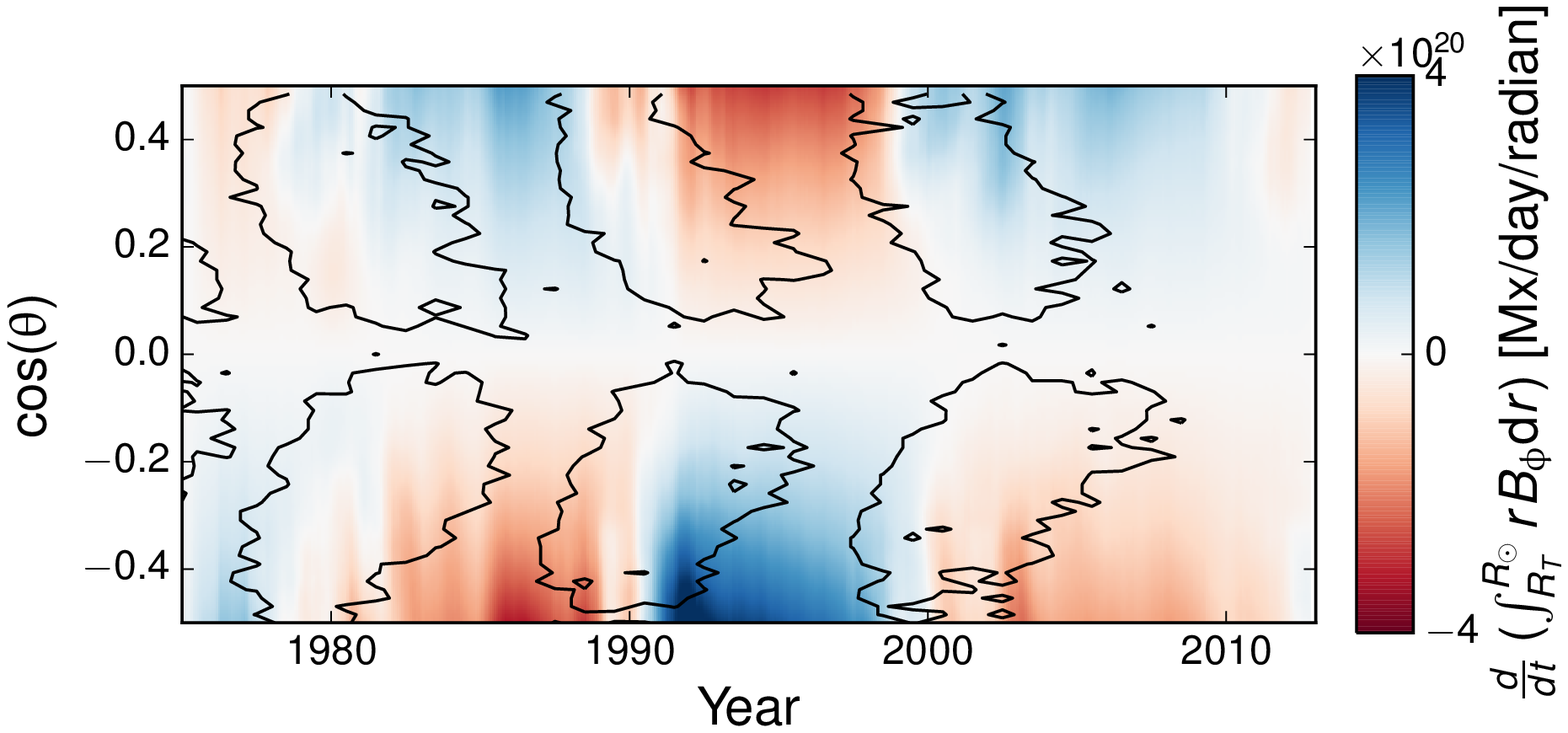}  
\includegraphics[scale=0.33,clip = true]{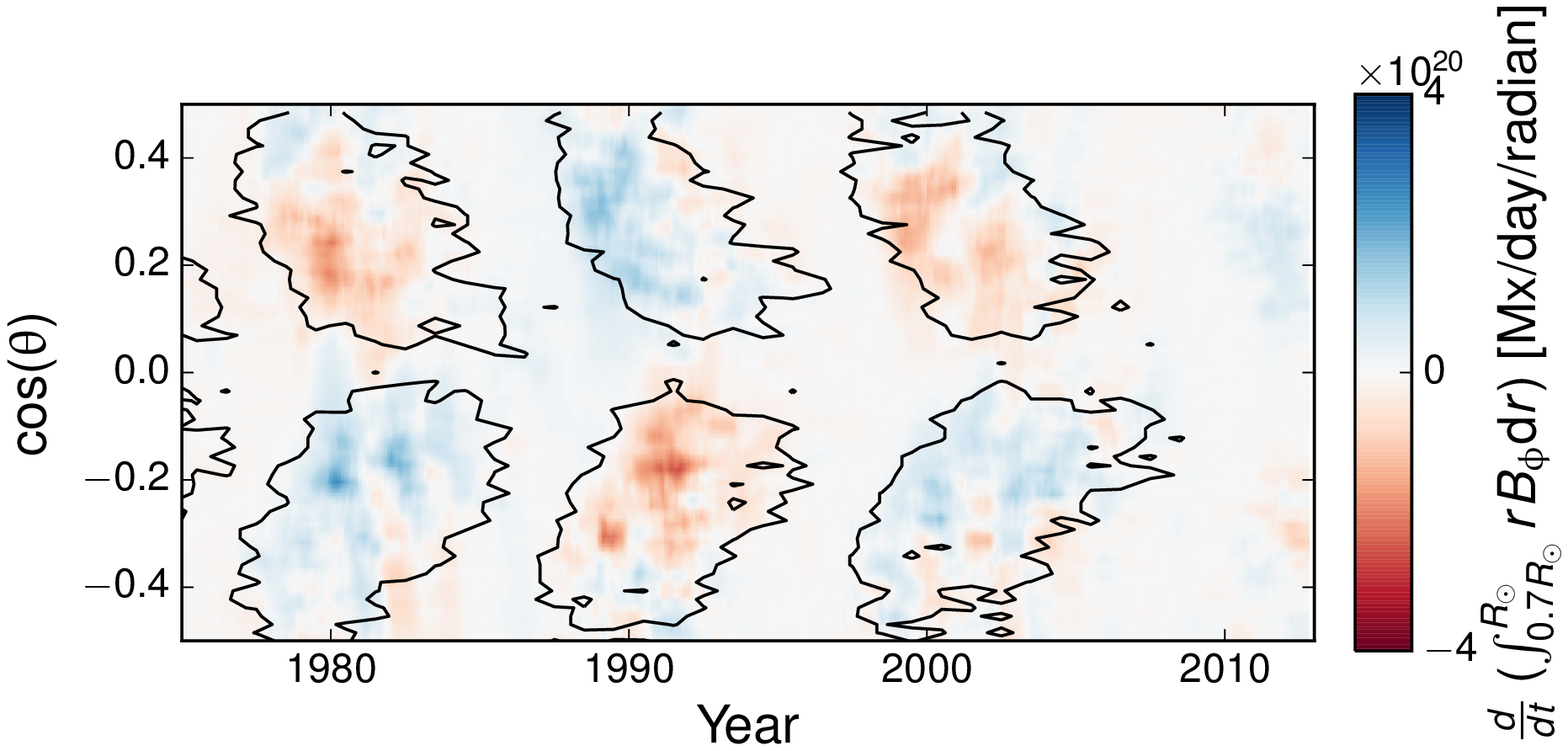}

\includegraphics[scale=0.33,clip = true]{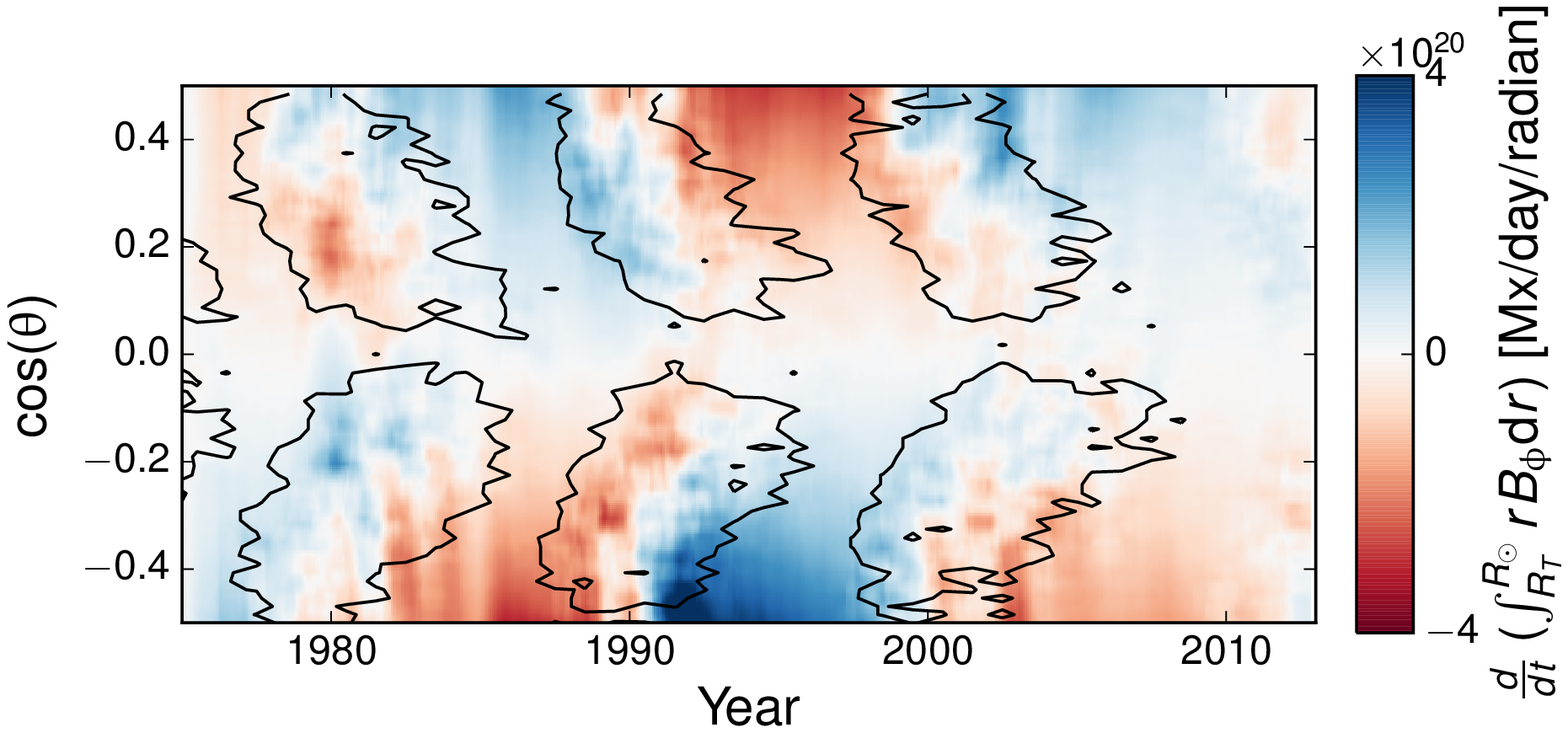}
\includegraphics[scale=0.33,clip = true]{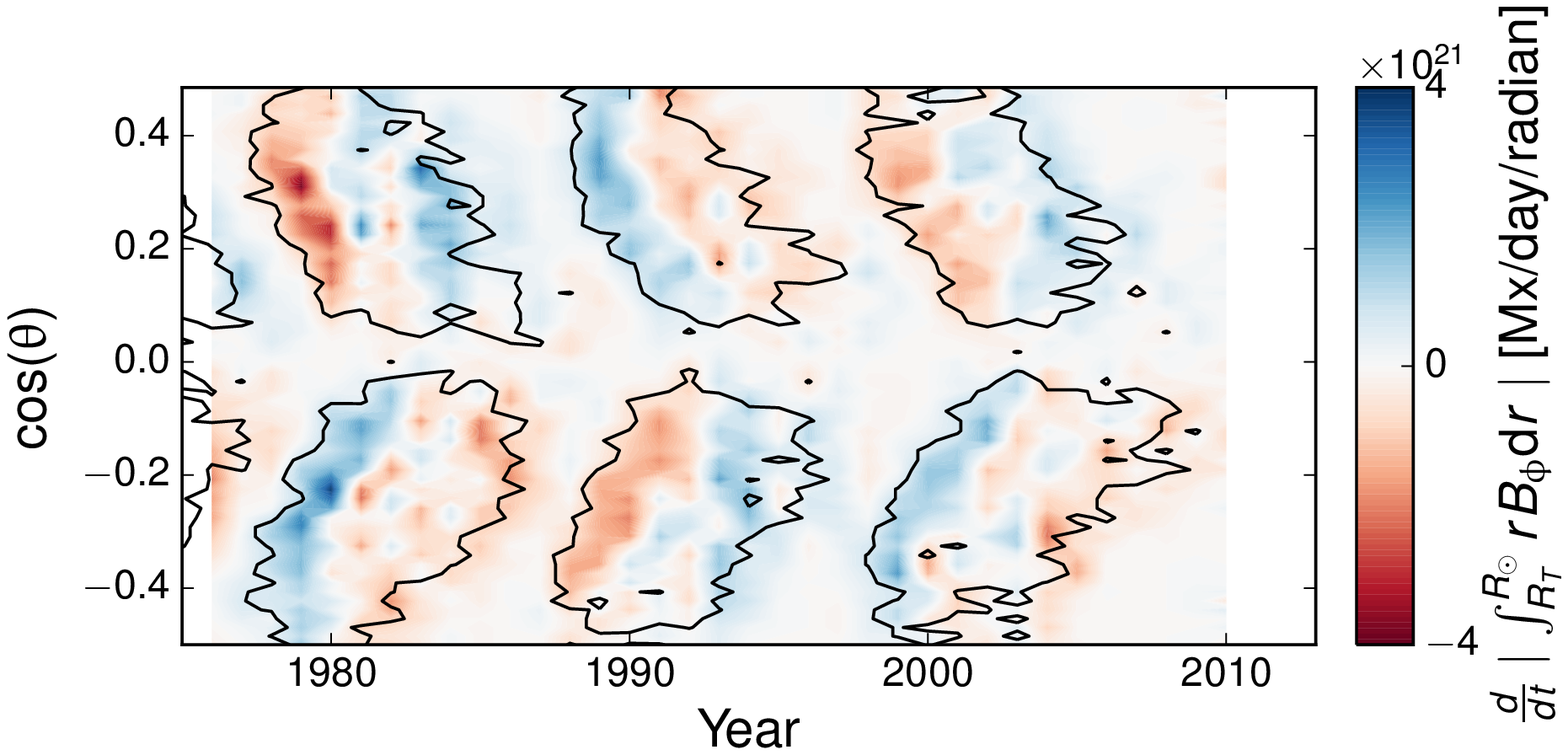}
    
\caption{Generation rate of radially integrated toroidal magnetic flux
         as a function of time and latitude based on KPNO/VTT and SOLIS
         synoptic magnetograms. {\em Upper left}: Flux generation by
         latitudinal shear.  {\em Upper right}: Flux generation by
         radial shear.  {\em Lower left}: Flux generation by latitudinal
         and radial shear combined. {\em Lower right}: Change of
         integrated toroidal flux estimated from the sunspot emergence
         rate. Note the different scaling for the colour bar of this
         plot.  Black contours outline the region where sunspots were
         observed to emerge.}
  \label{fig:KPNO}
\end{figure*}

\end{appendix}

\end{document}